\documentclass{article}
\usepackage{spconf}
\usepackage[T1]{fontenc}
\usepackage[latin1]{inputenc}
\usepackage{graphicx}
\usepackage{subfigure}
\usepackage{color}
\usepackage{mathrsfs}
\usepackage{amsfonts}
\usepackage{amsmath}
\newcommand{\avg}[1]{\left< #1 \right>} % for average

\newtheorem{mydef}{Definition}[section]

\begin{document}

% Example definitions.
% --------------------
\def\x{{\mathbf x}}
\def\L{{\cal L}}

% Title.
% ------
%\title{Strategy to manage microgrids using the model-free control approach }
%
\title{Model-free control of nonlinear power converters  }
% Single address.
% ---------------
\name{Loïc Michel$^{\ast}$, Wim Michiels$^{\ast}$ and Xavier Boucher$^{\dag}$}

\address{$^{\ast}$ Department of Computer Science \\
KU Leuven \\
Celestijnenlaan 200A \\
B - 3001 Heverlee \\
E-mail: $\{$Loic.Michel, Wim.Michiels$\}$@cs.kuleuven.be \\
$^{\dag}$ E-mail: xavier.b.eng@gmail.com}
%
% For example:
% ------------
%\address{School\\
%	Department\\
%	Address}
%
% Two addresses (uncomment and modify for two-address case).
% ----------------------------------------------------------
%\twoauthors
%  {A. Author-one, B. Author-two\sthanks{Thanks to XYZ agency for funding.}}
%	{School A-B\\
%	Department A-B\\
%	Address A-B}
%  {C. Author-three, D. Author-four\sthanks{The fourth author performed the work
%	while at ...}}
%	{School C-D\\
%	Department C-D\\
%	Address C-D}
%

%\ninept
%
\maketitle
\begin{abstract}
A new "model-free" control methodology is applied to a boost power converter. The properties of the boost converter allow to evaluate the performances of the model-free strategy in the case of switching nonlinear transfer functions, regarding load variations. Our approach, which utilizes "intelligent" PI controllers, does not require any converter model identification while ensuring the stability and the robustness of the controlled system. Simulation results show that, with a simple control structure, the proposed control method is almost insensitive to fluctuations and large load variations.
\end{abstract}
\section{Introduction}
\label{sec:intro}
The model-free control methodology, originally proposed by \cite{esta}, has been widely successfully applied to many mechanical and electrical processes. The model-free control provides good performances in disturbances rejection and an efficient robustness to the process internal changes. A preliminary work on power electronics \cite{Michel} presents the successful application of the model-free control method to the control of dc/dc converters. The control of nonlinear power converters has been deeply studied and some advanced methods have been successfully developed and tested (e.g. \cite{Sira} \cite{jp}  \cite{Siew} \cite{Morroni_2} \cite{Galotto}).  This paper extends the previous results to the control of the boost converter working in different conduction modes. In particular, we will show that the proposed control method is robust to strong load changes that may involve switching working modes.

The paper is structured as follows. Section II presents an overview of the model-free control methodology including its advantages in comparison with classical methodologies. Section III presents the basic theory of the boost converter. Section IV discusses the application of the model-free control to the boost converter. Some concluding remarks may be found in Section V.

\section{Model-free control: a brief overview}\label{mfc}
\subsection{General principles}

We only assume that the plant behavior is well approximated in its operational range by a system of ordinary differential equations, which might be highly nonlinear and time-varying. The system, which is SISO, may be therefore described by the input-output equation:

\begin{equation}\label{es}
E (t, y, \dot{y}, \dots, y^{(\iota)}, u, \dot{u}, \dots, u^{(\kappa)}) = 0
\end{equation}

\begin{itemize}
\item $u$ and $y$ are the input and output variables,
\item $E$, which might be unknown, is assumed to be a
sufficiently smooth function of its arguments.
\end{itemize}

From (\ref{es}), we define an {\it ultra-local} model, which represents (\ref{es}) over a small time period.

\begin{mydef}\label{mydef-modele_F}
\cite{esta} If $u$ and $y$ are respectively the variables of input and output of a system to be controlled, then this system can be described as the ultra-local model defined by:
\begin{equation}\label{mod}
y^{(n)} = F + \alpha u
\end{equation}
where
%\begin{itemize}
%\item
$\alpha \in \mathbb{R}$ is a {\em non-physical} constant parameter,
such that $F$ and $\alpha u$ are of the same magnitude,
and $F$ contains all structural information of the process.
%\item the numerical value of $F$, which contains the whole ``structural information'',
%is determined from $u$, $\alpha$, and an
%estimate of the derivative $y^{(n)}$.
%\end{itemize}
\end{mydef}
In all the numerous known examples, it was possible to set $n = 1$ or $2$ \cite{Fliess_Mar}. Let us emphasize that one only needs to give an approximate numerical value to $\alpha$. The gained experience shows that taking $n = 2$ allows to stabilize switching systems.

\subsection{Intelligent PI controllers}

%    premier_ordre = 2;  % ordre de la dérivation sur la consigne et la sortie
%    alpha = 30;    % réglage de alpha pour le sane-modèle
%    CSM = 0;       % correction PI : 0 - correction sans-modèle : 1 - pas de correcteur : 2
%    sat = 0;       % pas de saturation sur le duty cycle : 0
%    power = 0;     % contrôle en puissance : 0 - contrôle en tension 1 - contrôle en puissance
%    gain = 1;   % selection correcteur PI ou gain pur
%    d_div = 200;   % facteur de division sur la sortie
%
%Kp = 0.5;       % proportionnal integrator
%Ki = 0;       % integrator coefficient

\begin{mydef}
\cite{esta} \label{mydef-iPI} We close the loop via the {\em intelligent PI controller}, or {\em i-PI} controller,
\begin{equation}\label{eq:ipi}
u = - \frac{[F]}{\alpha} + \frac{{y}^{(n) \, \ast}}{\alpha}  + \mathcal{C}(\varepsilon)
\end{equation}
where
\begin{itemize}
\item $[F]$ is an estimate of $F$ in (\ref{mod}), computed on-line as $[y^{(n)}]-\alpha u$, where $[y^{(n)}]$ is an approximation of the output derivative;
\item $y$ is the measured output to control and $y^\ast$ is the output reference trajectory;
\item $\varepsilon = y^\ast - y$ is the tracking error;
\item $\mathcal{C}(\varepsilon)$ is of the form $K_p \varepsilon + K_i \int \varepsilon$. $K_p$, $K_i$ are the usual tuning gains.
\end{itemize}

Equation (\ref{eq:ipi}) is called the model-free control law or model-free law.
\end{mydef}

%%%%%%%%%%%%%%%%%%%%%%%%%%

The i-PI controller (\ref{eq:ipi}) is compensating the poorly known term $F$ and controlling the system therefore boils down to the control of an integrator. The tuning of the gain $K_p$ and $K_i$ becomes therefore straightforward.

Our implementation of (\ref{eq:ipi}) assumes a sampled-data control context, where the control input is kept constant over the inter-sampling interval and the output derivatives are approximated by finite-differences of the outputs. At the $k$th sampling instants, we have \cite{Michel}:
\begin{multline}\label{eq:ipi2}
%\begin{array}{lr}
u_k = u_{k-1} - \frac{1}{\alpha T_c^2} \left\{ \left( y_{k-1} - 2 y_{k-2} + y_{k-3} \right) \right. -   \\
\left. \left( y^*_{k-1} - 2 y^*_{k-2} + y^*_{k-3} \right) \right\} + C(y_{k-1}^*-y_{k-1})
%\end{array}
\end{multline}

\noindent
where $u_k$ refers to the averaged duty-cycle at the $k$th sampling instant and $T_c = 0.1$ ms is the switching period. The main advantage of the proposed control approach is that sudden changes in the model, e.g. due to load changes, and model uncertainty can be overcome as $F$ in (\ref{mod}) is re-estimated at every sampling instant from the output and inputs derivatives. We note that the potential amplification of noise by differentiation of the output can be countered by using moving average filters, see \cite{fliess}.

%Since $F$, which represents the model of the plant during a small period, is estimated and updated at each sample $k$, it means that the model-free control would take into account any changes / disturbances of the process and in particular, load changes and components uncertainties.

%To illustrate the utilization of the model-free control applied to the control of the boost conveter, the following results present the simulation of a voltage-controlled inverter, a three-phase controlled inverter and a power controlled inverter under disturbances such as (e.g.) load changes. We compare the results with a PI control that has been tuned using an ITAE criteria in order to optimize the transient with the initial load \cite{Awouda}. Simulations have been performed using the averaging method \cite{Sun} \cite{sigma} for which the controlled inputs in every case correspond to the averaged duty-cycle values that drive each IGBT.
\section{The boost converter}

\begin{figure}[!h]
\centering
\includegraphics[width=8.5cm]{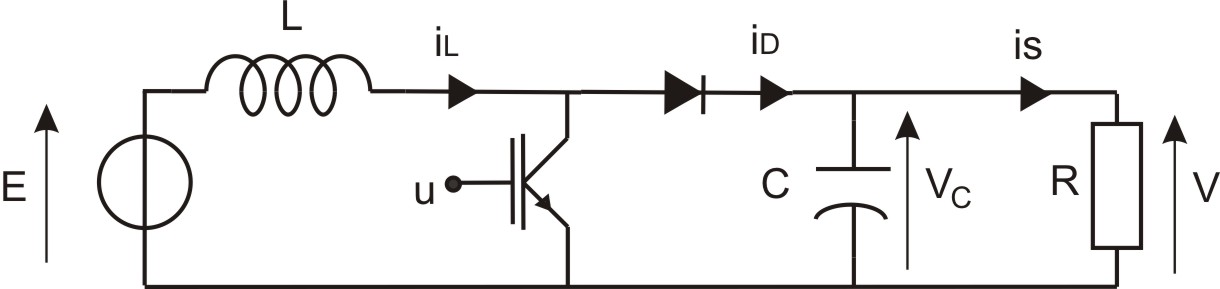}
\caption{Boost converter.}
\label{fig:these_boost}
\end{figure}

The boost converter is a well-known power converter that produces a dc output voltage greater in magnitude than the dc input voltage \cite{Erickson}. It has two working modes, the continuous mode and the discontinuous mode, that involve two different nonlinear transfer functions according to the output current. Consider the boost converter depicted in Fig. \ref{fig:these_boost} whose parameters are given in the Tab. \ref{table:Boost}. The parameter $f_c$ is the switching frequency.

\begin{table}[!b]
\renewcommand{\arraystretch}{1.3}
\centering
\caption{\label{table:Boost} Parameters of the boost converter.}
\begin{tabular}{c||c}
\hline
\bfseries Composant & \bfseries Value \\
\hline\hline
$L$ & 10 mH\\
$C$ & 47 $\mu$F\\
$R$ & 50 - 200 $\Omega$ \\
$f_c$ & 10 kHz \\
\hline
\end{tabular}
\end{table}

A full state-space averaged model, derived by \cite{JSun}, allows to describe the working modes of the boost converter. The model-free control implementation is based on this formulation and aims to control the output voltage $V$ according to a reference output $V^*$ .

\subsection{Nonlinear average modeling}

There exist three possible states according to the value of the duty-cycle. Denote $u$ the instantaneous control variable of the boost converter. $d_1$ defines the on-state duration (i.e. $u = 1$), $d_2$, the off-state duration in the continuous mode (i.e. $u = 0$ and $i_L \neq 0$) and $d_3$, the off-state duration in the  discontinuous mode (i.e. $u = 0$ and $i_L = 0$). The equations take the form:

\begin{equation}\label{eq:boost_disc_SS}
\begin{array}{l}
\dot{x} = A_1 x + b_1 E \hbox{ for } t \in [0, \, d_1 T_s] \\
\dot{x} = A_2 x + b_2 E \hbox{ for } t \in [d_1 T_s, \, (d_1 + d_2) T_s] \\
\dot{x} = A_3 x + b_3 E \hbox{ for } t \in [(d_1 + d_2)T_s, \, T_s] \\
\end{array}
\end{equation}

\noindent
where $(A_1, b_1)$, $(A_2, b_2)$ et $(A_3, b_3)$ represent respectively the state-spaces of conduction, non-conduction and discontinuous conduction.
%\noindent
%The state-space averaged model associated to the system of equations (\ref{eq:boost_disc_SS}) verifies :
%
%\begin{equation}\label{eq:boost_disc_SS_2}
%\begin{array}{l}
%\avg{\dot{x}} = [d_1 A_1 + d_2 A_2 + (1 - d_1 - d_2) A_3] \avg{{x}}   \\
%+[d_1 b_1 + d_2 b_2 + (1 - d_1 - d_2) b_3 ] E  \\
%\end{array}
%\end{equation}
%
%\noindent
%where:

\noindent
with:

\begin{equation}
A_1 = \left( \begin{array}{cc}
            0 & 0 \\
            0 & -\frac{1}{RC}
          \end{array} \right) \,
A_2 = \left( \begin{array}{cc}
            0 & \frac{1}{L} \\
            \frac{1}{C} & -\frac{1}{RC}
           \end{array} \right)
           \end{equation}
\begin{equation}
A_3 = \left( \begin{array}{cc}
            0 & 0 \\
            0 & -\frac{1}{RC}
          \end{array} \right)
\end{equation}

\begin{equation}
b_1 = \left( \begin{array}{c}
            \frac{1}{L} \\
            0
          \end{array} \right) \qquad
b_2 = \left( \begin{array}{c}
            \frac{1}{L} \\
            0
          \end{array} \right) \qquad
b_3 = \left( \begin{array}{c}
            0 \\
            0
          \end{array} \right)
\end{equation}

\noindent
Ref. \cite{JSun} presents a derivation of (\ref{eq:boost_disc_SS}) in order to obtain a single state-space averaged representation that models the three working modes depending on the value of $d_1$. Assuming that the state vector of the boost converter (Fig. \ref{fig:these_boost}) is written $x = (i_L \, v_C)^T$, the state-space representation of the boost converter reads:

\begin{equation}\label{eq:boost_full_SS}
            \left( \begin{array}{c}
            \avg{\dot{i_L}} \\
            \avg{\dot{v_C}}
           \end{array} \right) =
          \left( \begin{array}{cc}
            0 & -\frac{d_2}{L} \\
            \frac{d_2}{C} & -\frac{1}{RC}
           \end{array} \right)
          M
          \left( \begin{array}{c}
            \avg{{i_L}} \\
            \avg{{v_C}}
           \end{array} \right) +
         \left( \begin{array}{c}
            \frac{d_1 + d_2}{L} \\
            0
           \end{array} \right) E
\end{equation}

\noindent
where $M$ is a matrix that corrects the resulting state-space according to the number of reactive elements in the circuit \cite{Chiniforoosh}.

\begin{equation}
M = \left( \begin{array}{cc}
            \frac{1}{d_1 + d_2} & 0 \\
            0 & 1
          \end{array} \right)
\end{equation}

\noindent
The input of this state-space model $d_1$ is defined by $d_1 \equiv \avg{u}$. The averaged output voltage $V = \avg{v_C}$ from (\ref{eq:boost_full_SS}) is directly used as the feedback measurement to compute the model-free control.

\subsection{Conduction modes}

We define two conduction modes of the boost converter according to the $i_L$ current. In particular,
\begin{itemize}
\item {\it The continuous mode (CCM)} is defined by a $i_L$ current which does not vanish during a switching period. The static transfer relation between the input $u$ and the output $V$ verifies:
    \begin{equation}
    \frac{V}{E} = \frac{1}{1-\avg{u}}
    \end{equation}
The relation between the on-state $d_1$ and the off-state $d_2$ verifies:
\begin{equation}
d_2 = f(d_1) = 1 - d_1
\end{equation}
\item {\it The discontinuous mode (DCM)} is defined by a $i_L$ current which vanishes during a switching period. The static transfer relation between the input $u$ and the output $V$ verifies:
    \begin{equation}
    \avg{v_{C}} = E \left[ 1 + \avg{u^2} \frac{E}{2 L f_c \avg{i_s}} \right]
    \end{equation}
The relation between the on-state $d_1$ and the off-state $d_2$ verifies:
    \begin{equation}
    d_2 = f(d_1, \avg{v_C}, E) = \frac{E}{\avg{v_C} - E} d_1
    \end{equation}
\end{itemize}

According to the value of the duty-cycle $\alpha$ and the maximum magnitude of the current going through the diode $|I_D|$, the passage from the CCM and DCM modes can be described by a nonlinear relation such as:

\begin{equation}\label{eq:mode_switch}
\frac{1}{2 |I_D|} d_1 ( 1 - d_1) = 1
\end{equation}

\noindent
Note that in the DCM, $d_2$ depends not only of $d_1$ but also on $\avg{v_C}$, which results finally in a nonlinear transfer function \cite{Erickson}.

\section{Simulation results}

\begin{figure}[!h]
\centering
\includegraphics[width=8.5cm]{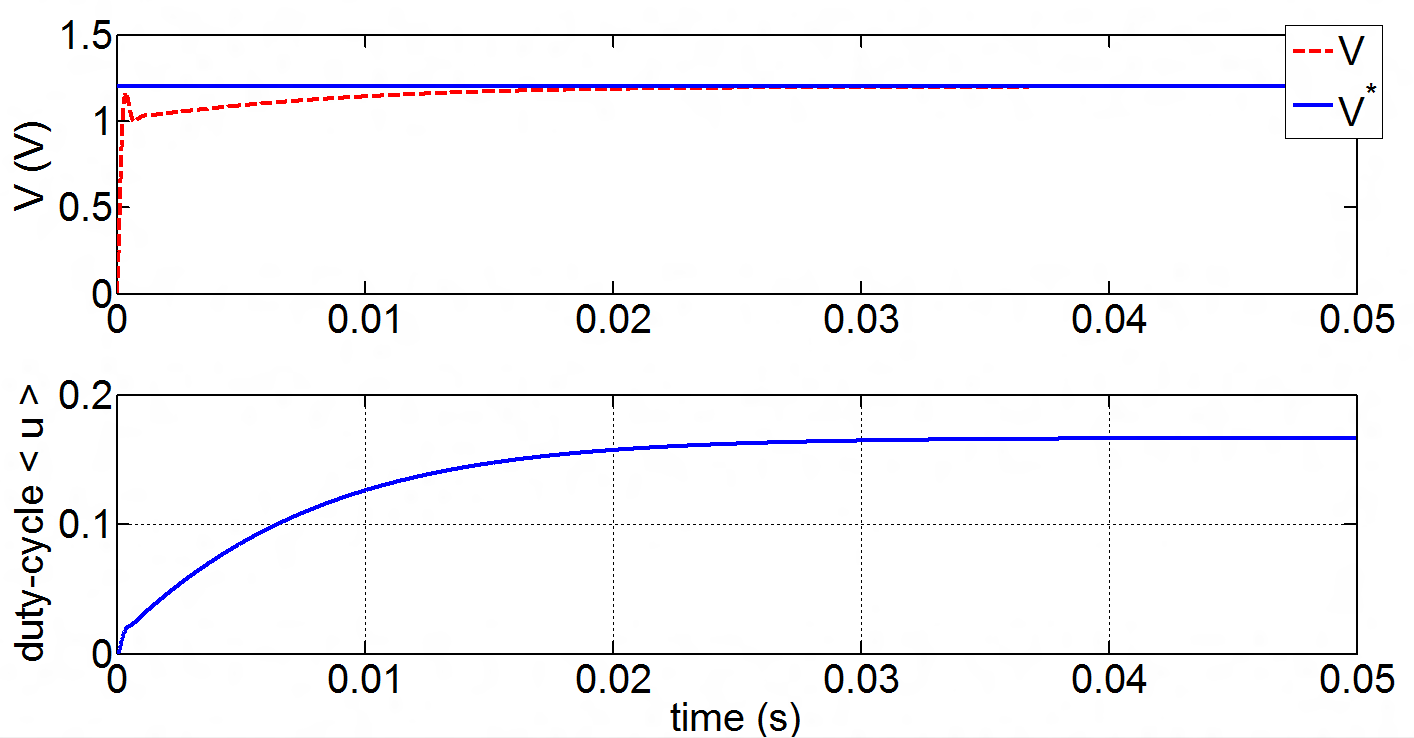}
\caption{Tracking and stabilization with a constant output reference.}
\label{fig:fig_boost_CSM_1}
\end{figure}

We adjust the parameters of the model-free control (\ref{eq:ipi2}), which are not correlated to the model, such as the boost converter is stabilized at the beginning of the simulation, whatever the initial conduction mode. In particular, we take $K_p = 2$, $K_i = 10$ and $\alpha = 30$. Figure \ref{fig:fig_boost_CSM_1} presents the stabilization of the output voltage $V$ following a constant output reference $V^*$.

\begin{figure}[!h]
\centering
\includegraphics[width=8.5cm]{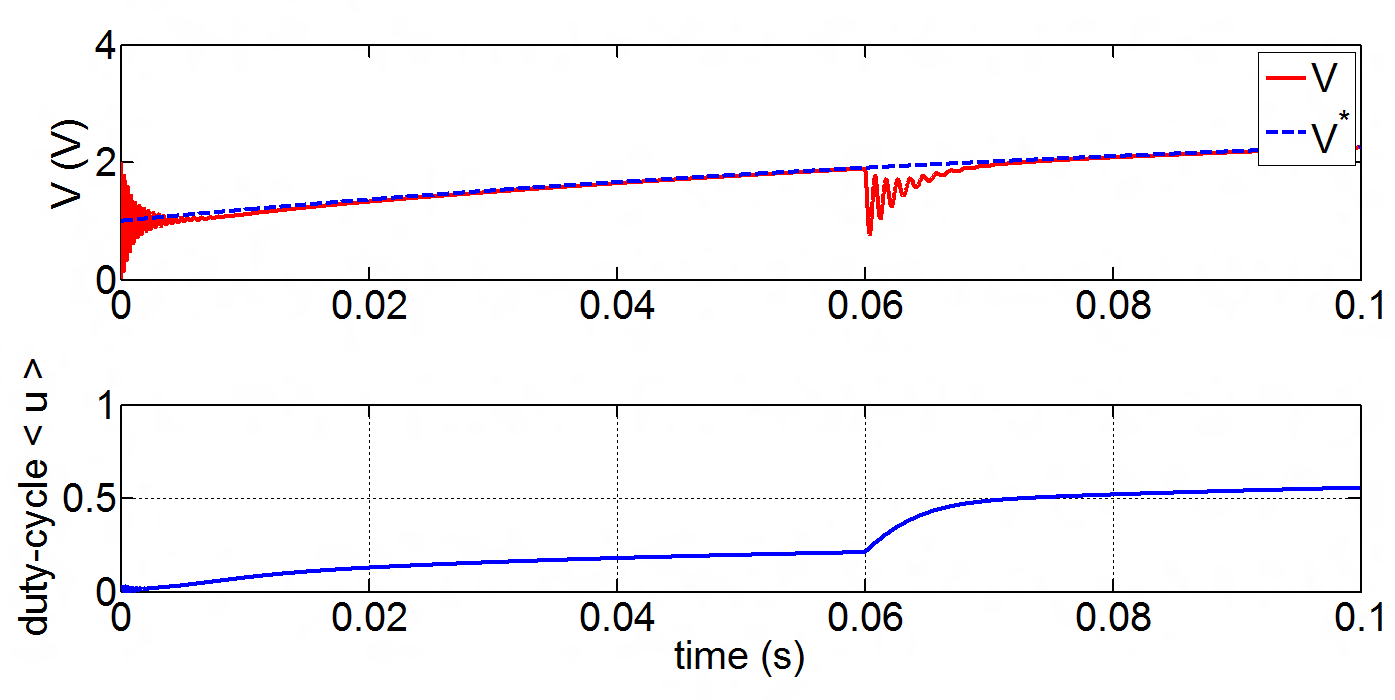}
\caption{Tracking of an exponential output reference - at t = 0.06 s, the load switches from $R = 100 \, \Omega$ to $R = 50 \, \Omega$, inducing a switch from
DCM to CCM.}
\label{fig:fig_boost_CSM_2}
\end{figure}

\begin{figure}[!h]
\centering
\includegraphics[width=8.5cm]{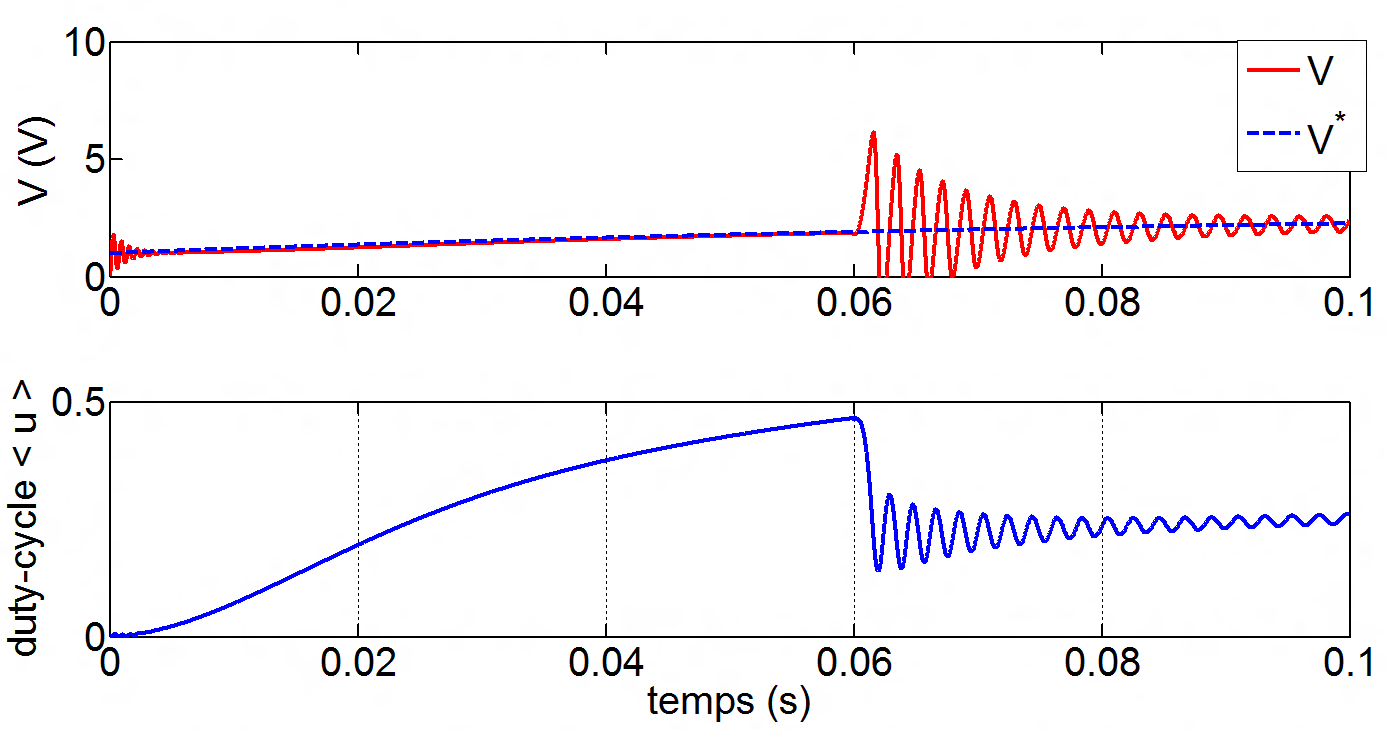}
\caption{Tracking of an exponential output reference - at t = 0.06 s, the load switches from $R = 60 \, \Omega$ to $R = 100 \, \Omega$, inducing a switch from
CCM to DCM.}
\label{fig:fig_boost_CSM_3}
\end{figure}

\begin{figure}[!h]
\centering
\includegraphics[width=8.5cm]{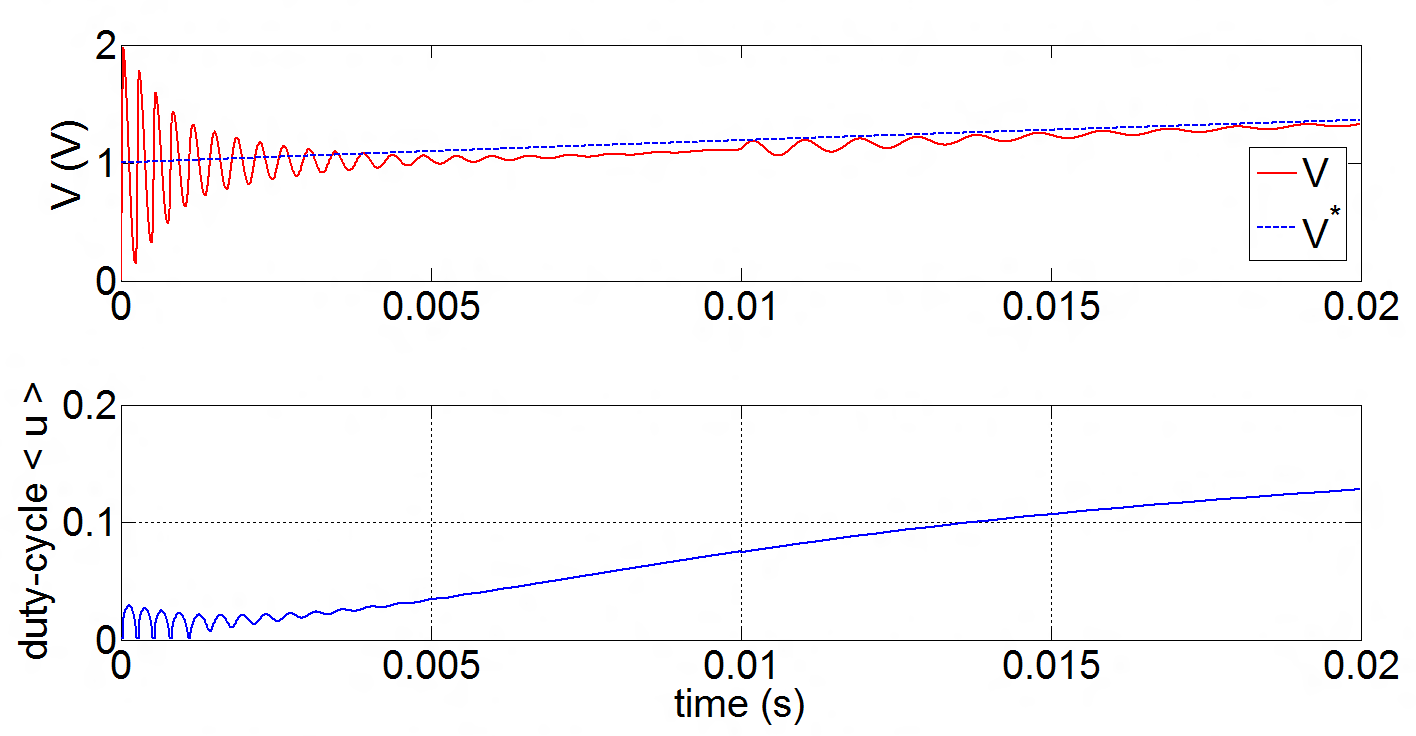}
\caption{Tracking of an exponential output reference under the DCM with a load change from $R = 100 \, \Omega$ to $R = 200 \, \Omega$ at $t = 0.01$ s.}
\label{fig:fig_boost_CSM_4}
\end{figure}

Figures \ref{fig:fig_boost_CSM_2} and \ref{fig:fig_boost_CSM_3} present the tracking of an exponential output reference $V^*$ in presence of a load change at $t = 0.06$ s that induces a switch of the working mode, according to (\ref{eq:mode_switch}). Figure \ref{fig:fig_boost_CSM_4} focuses on the initial resonant transient in the CCM in presence of a load change that does not involve a switch of the working mode.

\noindent
We observe that the model-free control is able to stabilize the output voltage even if the change of the load induces a switch between the two conduction modes. To damp the oscillations that occur during each switching transient, an advanced model-free controller can be used \cite{sympl}.

\section{Concluding remarks}

The simulations on the boost power converter aims to generalize the use of the model-free control to nonlinear power systems. Simulations show that the model-free control methodology yields robust performances with respect to disturbance rejection as well as the stabilization of the switching modes. This methodology moreover does not need any well-defined mathematical model, that would require some complex identification procedure. A single parameter only, $\alpha$, needs to be tuned properly in order to ensure a large variety of working points.

\section*{Acknowledgements} The article presents results of the project G.0717.11 of the Research Council Flanders (FWO).

%List and number all bibliographical references at the end of the
%paper.  The references can be numbered in alphabetic order or in order
%of appearance in the document.  When referring to them in the text,
%type the corresponding reference number in square brackets as shown at
%the end of this sentence \cite{C2}.
%
%% References should be produced using the bibtex program from suitable
%% BiBTeX files (here: strings, refs, manuals). The IEEEbib.bst bibliography
%% style file from IEEE produces unsorted bibliography list.
%% -------------------------------------------------------------------------
%\bibliographystyle{IEEEbib}
%\bibliography{strings,refs}

\begin{thebibliography}{}

\bibitem{esta} M. Fliess and C. Join, ``Commande sans mod\`{e}le et commande \`{a} mod\`{e}le
restreint``, e-STA, vol. 5 (n$^\circ$ 4), pp. 1-23, 2008. %(available
%at \url{http://hal.inria.fr/inria-00288107/en/}).

\bibitem{Michel} L. Michel, C. Join, M. Fliess, P. Sicard and A. Chériti, ``Model-free control of dc/dc converter``, in 2010 IEEE 12th Workshop on Control and Modeling for Power Electronics, pp.1-8, June 2010. %(available at \url{http://hal.inria.fr/inria-00495776/}).

\bibitem{Sira} H. Sira-Ramirez and R. Silva-Ortigoza, ``Control Design Techniques in Power Electronics  Devices``, Springer, 2009.

\bibitem{jp} H. Fujioka, C.-Y. Kao, S. Alm\'er and U. J\"onsson, ``Robust tracking
with {\bf H}$_\infty$ performance for PWM systems``, Automatica, vol. 45, pp. 1808-1818, 2009.

\bibitem{Siew} Tan Siew-Chong, Y.M. Lai, M.K.H. Cheung and C.K. Tse, ``On the practical design of a sliding mode voltage controlled buck converter``, IEEE Trans. Power Electron., vol. 20, pp. 425-437, 2005.

\bibitem{Morroni_2} J. Morroni, L. Corradini, R. Zane and D. Maksimovic, ``Adaptive tuning of switched-mode power supplies operating in discontinuous and continuous conduction modes``, IEEE Trans. Power Electron., vol. 24, pp.2603-2611, 2009.

\bibitem{Galotto} L. Galotto,  C.A. Canesin, R.  Cordero, C.A.  Quevedo and R.  Gazineu, ``Non-linear controller applied to boost DC-DC converters using the state space average model``, in COBEP '09. Brazilian Power Electronics Conference, pp.733-740, Oct. 2009.


\bibitem{Fliess_Mar} M. Fliess, C. Join, and S. Riachy, ``Nothing is as Practical as a Good Theory: Model-Free Control``, 4èmes Journées Doctorales / Journées Nationales, MACS, JD-JN-MACS, June 2011.


\bibitem{fliess} M. Fliess, C. Join and H. Sira-Ramirez, ``Non-linear estimation is easy``, Int. J. Model. Identif. Control, vol. 4, pp. 12-27, 2008. % (available at
%http://hal.inria.fr/inria-00158855/en/).

\bibitem{Erickson} R. Erikson and D. Maksimovic, ``Fundamental of power electronics``, Kluwer Academics Publishers, 2001.

\bibitem{JSun} J. Sun, D.M. Mitchell,  M.F. Greuel, P.T. Krein and R.M. Bass, ``Averaged modeling of PWM converters operating in discontinuous conduction mode``, IEEE Transactions on  Power Electronics, vol.16, no.4, pp.482-492, Jul. 2001.

\bibitem{Chiniforoosh} S. Chiniforoosh, J. Jatskevich, A. Yazdani, V.  Sood, V. Dinavahi, J.A. Martinez and A.  Ramirez,  ``Definitions and Applications of Dynamic Average Models for Analysis of Power Systems``, IEEE Transactions on Power Delivery, vol.25, no.4, pp.2655-2669, Oct. 2010.

\bibitem{sympl} L. Michel, ``Variational and symplectic approach of the model-free control``, preprint ArXiv, 2010.

\end{thebibliography}

\end{document}